\documentclass{aastex}
\usepackage{emulateapj5}

\shorttitle{ Imprint of Gravitational Lensing
}
\shortauthors{Hirose et al.}

\begin{document}


\title{Imprint of Gravitational Lensing by Population III Stars in 
Gamma Ray Burst Light Curves
}


\author{Y. Hirose\altaffilmark{1}, M. Umemura\altaffilmark{1}, 
A. Yonehara\altaffilmark{2,3}, and J. Sato\altaffilmark{1} }
\email{hirose@ccs.tsukuba.ac.jp}


\altaffiltext{1}{Center for Computational Sciences, University of Tsukuba, 
Ibaraki 305-8577, Japan }
\altaffiltext{2}{Department of Physics, University of Tokyo, 
Hongo, Bunkyo, Tokyo 113-0033, Japan}
\altaffiltext{3}{Astronomisches Rechen-Institut, Zentrum f\"{u}r Astronomie, 
Universit\"{a}t Heidelberg, M\"{o}nchhofstra{\ss}e 12-14, 
69120 Heidelberg, Germany}


\begin{abstract}
We propose a novel method to extract the imprint of gravitational 
lensing by Pop III stars in the light curves of 
Gamma Ray Bursts (GRBs). 
Significant portions of GRBs can originate in hypernovae of Pop III stars
and be gravitationally lensed by foreground Pop III stars or their remnants.
If the lens mass is on the order of $10^2-10^3M_\odot$ and the lens redshift
is greater than 10, the time delay between two lensed images of a GRB
is $\approx 1$s and the image separation is $\approx 10\ \mu$as. 
Although it is difficult to resolve the two lensed images spatially
with current facilities, the light curves of two images 
are superimposed with a delay of $\approx 1$ s. 
GRB light curves usually exhibit noticeable variability, where 
each spike is less than 1s. If a GRB is lensed, all spikes 
are superimposed with the same time delay. 
Hence, if the autocorrelation of light curve with changing time interval
is calculated, it should show the resonance at the time delay of lensed
images. 
Applying this autocorrelation method to GRB light curves 
which are archived as the {\it BATSE} catalogue, 
we demonstrate that more than half of the light curves can
show the recognizable resonance, if they are lensed. 
Furthermore, in 1821 GRBs 
we actually find one candidate of GRB lensed by a Pop III star, 
which may be located at redshift $20-200$. 
The present method is quite straightforward and therefore provides 
an effective tool to search for Pop III stars at redshift greater than 10.
Using this method, we may find more candidates of GRBs lensed 
by Pop III stars in the data by the {\it Swift} satellite. 
\end{abstract}



\keywords{gamma-ray burst----gravitational lensing---Pop\rm{III} stars}

\section{Introduction}

The recent observation of the cosmic
microwave background by 
{\it Wilkinson Microwave Anisotropy Probe} ({\it WMAP})
suggests that the reionization
of the universe took place at redshifts of $8 \lesssim z \lesssim 14$
(Spergel et al. 2003; Kogut et al. 2003; Page et al. 2006; Spergel, D., et al. 2006). 
This result implies that first generation stars (Pop III stars) were possibly
born at $z \gtrsim 10$ if UV photons emitted from Pop
III stars are responsible for cosmic reionization. 
However, there is no direct evidence that Pop III stars actually formed
at $z> 10$. Obviously, it is impossible with current or near future
facilities to detect the emission from a Pop III star at such high redshifts
(e.g., Mizusawa et al. 2004). 
But, gamma ray bursts (GRBs) can be detected even at $z> 10$, 
if they arise there. GRBs are the only currently
available tool for probing first generation objects in the universe.
If one uses the data of absolute magnitude for GRBs with known redshifts, 
one can expect that more than half GRBs are detectable if they occur at $z> 10$ 
(Lamb $\&$ Reichart 2000). 
Roughly 3000 GRBs have been detected to date, but redshifts
have been measured only for 30 GRBs (Bloom et al. 2003),
among which the most distant is GRB~000131 
at $z=4.5$ (Andersen et al. 2000).  But, 
if empirical relations between the spectral properties and
the absolute magnitude are used, the GRBs detected to date may 
include events at $z>10$ (Lloyd-Ronning et al. 2002; Yonetoku et al. 2004;
Murakami et al. 2005). 
In addition, recently a new GRB satellite, {\it Swift}, has been
launched (Gehrels et al. 2004). {\it Swift} is now accumulating
more data of GRBs at a high rate. 
Recently, GRB~050904 detected by {\it Swift}, 
in terms of metal absorption lines and Lyman break, 
turns out to have occurred at $z=6.295$ (Kawai et al. 2006).

The discovery of the association between GRB~030329
and SN~2003dh has demonstrated that
at least a portion of long bursts in GRBs are caused 
by collapse of massive stars (Kawabata et
al. 2003; Price et al. 2003; Uemura et al. 2003). 
On the other hand, Pop III stars are expected to form in a top-heavy
fashion with the peak at $100-10^3M_\odot$ in the initial mass
function (IMF) (e.g., Nakamura \& Umemura 2001 and references therein). 
Also, the theoretical study by Heger \& Woosley (2002) suggests 
that Pop III stars 
between $100M_\odot$ and $140M_\odot$ may end
their lives as GRBs accompanied by the core collapse into black holes. 
Heger et al. (2003) estimate, 
assuming the IMF by Nakamura \& Umemura (2001),
that 5\% of Pop III stars can result
in GRBs. 
In the context of cold dark matter
cosmology, more than 10-30\% of GRBs are expected to 
occur at $z \gtrsim 10$, 
assuming that the redshift distributions of GRBs trace
the cosmic star formation history (Bromm \& Loeb 2002). 
Thus, observed GRBs highly probably contain GRB originating 
from Pop III stars at $z \gtrsim 10$.

The firm methods to measure redshifts are the detection of
absorption and/or emission lines of host galaxies of GRBs
(e.g., Metzger et al. 1997), or the Ly$\alpha$
absorption edge in afterglow (Andersen et al. 2000). 
However, these methods cannot be applied for all GRBs,
but have been successful to determine redshifts only for 30 GRBs 
(Bloom et al. 2003). 
Instead, some empirical laws have been applied to 
much more GRBs. They include a variability-luminosity relation 
(Fenimore \& Ramirez-Ruiz 2000), a lag-luminosity relation 
(Norris et al. 2000), $E_p$-luminosity relation (Amati et
al. 2002), and the spectral peak
energy-to-luminosity relation (Yonetoku et al. 2004). 
Applying these relations to GRBs, the redshift distributions of GRBs 
are derived
(Fenimore \& Ramirez-Ruiz 2000; Norris et al. 2000; 
Schaefer et al 2001; Lloyd-Ronning et al. 2002; Yonetoku et al. 2004). 
Some analyses conclude that a portion of GRBs are located 
at $z \gtrsim 10$. 
However, it is still controversial whether such an indirect 
technique is correct or not. 

In this paper, we propose a novel method to constrain
the redshifts of GRBs that may originate from Pop III stars 
at $z \gtrsim 10$. 
In the present method, the effects of the gravitational 
lensing by Pop III stars are considered. 
The lensing of GRBs is considered 
for the first time by Paczy\'nski (1986, 1987), who proposed the possibility
that a soft gamma-ray repeater is produced by
gravitational lensing of a single burst at cosmological distance.
Also, Loeb \& Perna (1998) first discussed
the microlensing effect of GRB afterglows, and Garnavich;
Loeb \& Stanek (2000) found the candidate microlensed afterglow (GRB~000301C).
The rates of such events are further discussed from theoretical
points of view (Koopmans \&
Wambsganss 2001; Wyithe \& Turner 2002; Baltz \& Hui
2005).
Blaes \& Webster
(1992) argue the method to detect cosmological clumped dark matter
by using the probability of detectable GRB lensing.
Nemiroff et al. (1993) and Marani et al. (1999)
search for the compact dark matter candidate using
actual GRBs data obtained by the Burst and Transient
Source Experiment ({\it BATSE}) satellite on the {\it
Compton Gamma Ray Observatory} satellite. 
They focus on large mass lenses up to $10^6{M_{\odot}}$, 
which cause the delay time-scale of several tens-100 s. 
On the other hand, Williams \& Wijers (1997) investigate
the influence on GRB light curve of the millisecond
gravitational lensing caused by each star in a lensing galaxy. 
In addition, Nemiroff \& Marani (1998) argue that 
it is possible to place constraints on the cosmic density of 
dark matter, baryons, stars, and so on, by microlensing
by stellar mass objects. In the present  method, we focus on the
gravitational lensing by Pop III stars. If the mass of
Pop III stars is on the order of $10^2-10^3M_\odot$ and
the redshift is greater than 10, the time delay between
two lensed images of a GRB is $\approx 1$ s. 
Quite advantageously, this time delay is longer than the time
resolution (64 ms) of GRB light curves and shorter than the
duration of GRB events, which is several tens to 100 sec
for long bursts. 
Thus, we can see the superimposed
light curves of two lensed images. 
The present method seeks for the imprint of gravitational 
lensing by Pop III Stars in GRB light curves.
We attempt to extract the imprint of lensing
by calculating the autocorrelation of light curves. 

In this paper, we assume a standard $\Lambda$CDM 
cosmological parameter: 
$H_0=70$ km $\rm{s}^{-1}$ $\rm{Mpc}^{-1}$, 
$\Omega_{\rm{M}}=0.3$, 
$\Omega_{\Lambda}=0.7$, and $\Omega_{\rm{b}}=0.04$.
The paper is organized as follows: In \S2, 
the formalism of gravitational lensing and the estimation of time delay 
between two images are provided. 
In \S3, the method to find the evidence of lensing by Pop III stars is proposed. 
Also, we demonstrate the potentiality of the present method 
for artificially lensed GRBs, and describe
how to determine the redshifts of lensed GRBs. 
In \S4, we apply this method to 1821 GRB data obtained by 
{\it BATSE}, and find a candidate of lensed GRB.
\S5 is devoted to the conclusions.

\section{Gravitational Lensing}

We consider a GRB lensed by a foreground Pop III star. 
Here, we presuppose the lens model of a point mass. 
The Einstein ring radius gives a typical scale of gravitational lensing, 
which  is expressed as
\begin{equation}
\theta_{\rm{E}} \equiv
\left( \frac{4GM_{\rm{L}}}{c^2}\frac{D_{\rm{LS}}}
{D_{\rm{OS}}D_{\rm{OL}}} \right)^{1/2},
\end{equation}
where $G$ is the gravity constant, $c$ is the speed of light,
$M_{\rm{L}}$ is the mass of a lens object, and
$D_{\rm{LS}}$, $D_{\rm{OS}}$, and $D_{\rm{OL}}$ are respectively
angular diameter distances between the lens and the
source, the observer and the source, and the observer and the
lens. A point mass lens produces two images with 
angular directions of
\begin{equation}
\displaystyle \theta=\frac{\beta}{2}\left[1\pm\sqrt{1+4\left(\frac{\theta_{\rm{E}}}{\beta}\right)^2}\right] \label{eq:theta} ,
\end{equation}
where $\beta$ is the angle of lens from a line-of-sight to the source. 
We hereafter express the image with $\theta>\theta_{\rm{E}}$ by image 1,
and that with $\theta<\theta_{\rm{E}}$ by image 2. 
The brightness of the image 1 and the image 2 
are respectively magnified by 
\begin{equation}
\displaystyle A_{1,2}=\frac{1}{4}[(1+4f^{-2})^{1/2}
+(1+4f^{-2})^{-1/2}\pm2], \label{eq:mag}
\end{equation}
where $f=\beta/\theta_{\rm E}$. Thus, the image 1 is
brighter than the original one, while the image 2 is fainter. 
In the case of a lens of Pop III star, the Einstein radius is estimated as
\begin{equation}
\displaystyle \theta_{\rm{E}}\simeq
10\left(\frac{M_{\rm{L}}}{10^3 M_{\odot}}\right)^{1/2}
\left(\frac{\tilde{D}}{4 \times 10^4 
\rm{Mpc}} \right)^{-1/2}\mu\rm{as} \label{eq:Einstein} ,
\end{equation}
where $\tilde{D} \equiv D_{OS} D_{OL} /D_{LS}$. 
Obviously, this angular separation is impossible to resolve by 
current facilities. 
Hence, we can just observe the superposition of two images.

\begin{figure} [p]
\begin{center}
\includegraphics[width=15cm,clip]{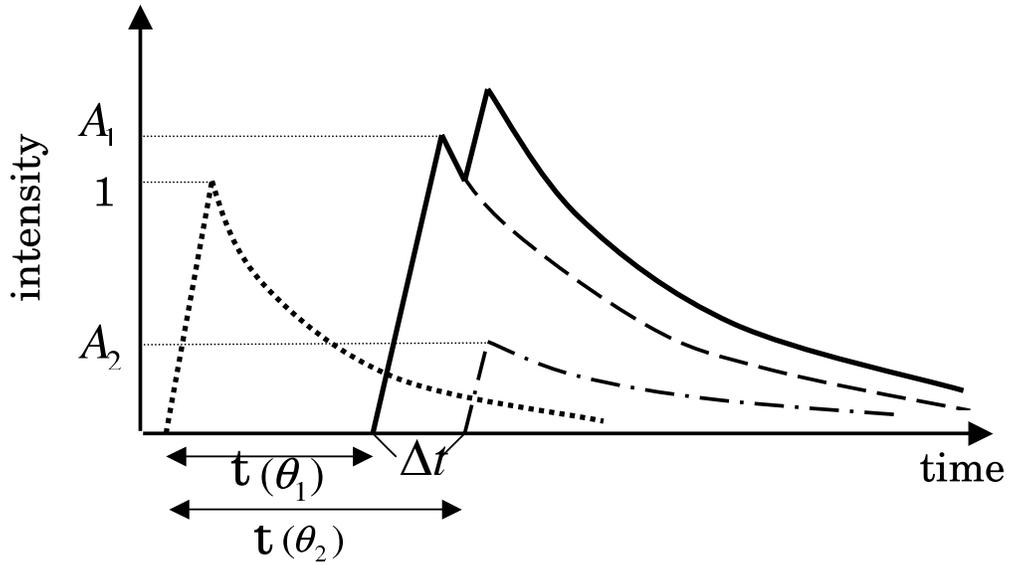}
\caption{Schematic diagram of lensed GRB light curve. The
dotted line represents the original light curve, while 
the dashed and dot-dashed lines represent the light curves of image 1 
and image 2, respectively. The superimposed light curve of lensed images
is shown by the solid line. 
The intensities are normalized by the maximum intensity of original light curve.}
\label{fig:image_of_lensed_lc}
\end{center}
\end{figure}

However, the light curves of two images are superimposed with
a time delay caused by the gravitational lensing, as shown in
Figure \ref{fig:image_of_lensed_lc}. 
The arrival time of signals for a lensed image is expressed as
\begin{equation}
\displaystyle t(\theta)=\frac{(1+z_{\rm{L}})}{c}\frac{D_{\rm{OS}}D_{\rm{OL}}}{D_{\rm{LS}}}\left[\frac{1}{2}(\theta-\beta)^2-\Psi(\theta)\right] \label{eq:delay} ,  
\end{equation}
where $z_{\rm L}$ is the redshift of lens object, 
and $\Psi$ is so called lens potential. 
For the point mass lens model, $\Psi$ is expressed as
\begin{equation}
\displaystyle \Psi(\theta)=\frac{D_{\rm{LS}}}{D_{\rm{OS}}D_{\rm{OL}}}\frac{4GM_{\rm{L}}}{c^2}\rm{ln}|\theta/\theta_{\rm{C}}| \label{eq:potential},
\end{equation}
where $\theta_{\rm{C}}$ is constant (Narayan \& Bartelmann 1997). 
Then, the time delay between two images is given by 
\begin{equation}
\Delta t(z_{\rm{L}},M_{\rm{L}},f)
=t(\theta_2)-t(\theta_1)\propto M_{\rm{L}}(1+z_{\rm{L}}) \label{eq:delta} .
\end{equation}
It should be noted that $\Delta t$ is determined solely by the mass and redshift 
of lens, regardless of the source redshift. 
In other words, the time delay places a constraint just on the lens, 
not on the source. 
However, if the lens redshift ($z_{\rm L}$) is determined, it gives 
the minimum value of the source redshift ($z_{\rm S}$) since 
$z_{\rm S}$ must be higher than $z_{\rm L}$. 

\begin{figure} [p]
\begin{center}
\includegraphics[width=15cm,clip]{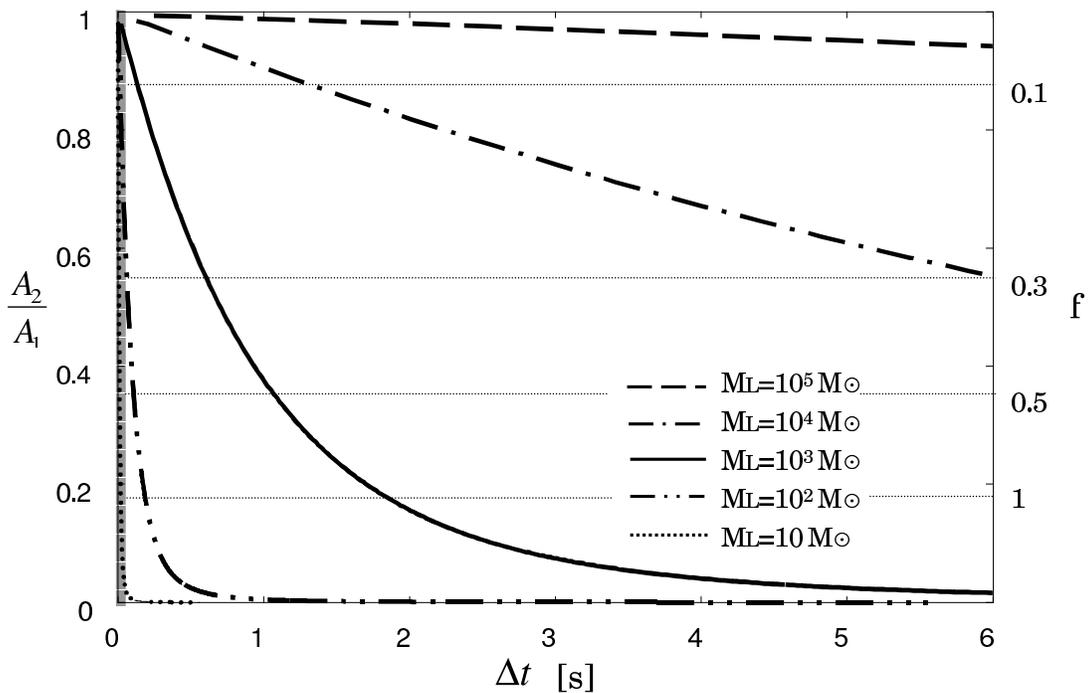}
\caption{Relation between
the time delay and the magnification ratio {\bf($A_2/A_1$) }
between two images. The redshift of lens object is fixed to 50. The
curves show lens masses of 10-$10^5M_{\rm{\odot}}$ with
an increment of 1 order of magnitude. 
In the right vertical axis, $f=\beta/\theta_{\rm E}$, is also shown
corresponding to the magnification ratio. 
The time resolution of {\it BATSE} data, $\Delta t = 64
\ \rm{ms}$,
is represented by the leftmost gray region.}
\label{fig:delay_amp}
\end{center}
\end{figure}

Fig. \ref{fig:delay_amp} illustrates the relation between
the time delay $\Delta t$ and the magnification ratio between 
image 1 and image 2, assuming the lens redshift of 50.
This figure shows that the lens with $\ga 10^4M_\odot$
yields the time delay longer than the standard GRB duration, 
if the typical delay time-scale is assessed by $f \approx 1$.
(Note that, as shown later, if $f$ becomes larger 
than unity, the ratio of magnification becomes 
smaller and therefore the contribution of image 2 becomes
difficult to extract. Also, if $f$ becomes smaller than unity, 
the probability of lensing goes down. ) 
On the other hand, the lens with $< 10M_\odot$ leads to 
$\Delta t$ shorter than the time resolution of light curves, 
and therefore the information of delay is buried. 
The mass scale of $10^2-10^3M_\odot$ expected for
Pop III stars gives $10^{-1} \ {\rm s} \la \Delta t \la 1 \ {\rm s} $,
which is longer than the time resolution and shorter than
the GRB duration. Hence, this mass range appears to be 
suitable for extracting the time delay information. 

\begin{figure} [p]
\begin{center}
\includegraphics[width=15cm,clip]{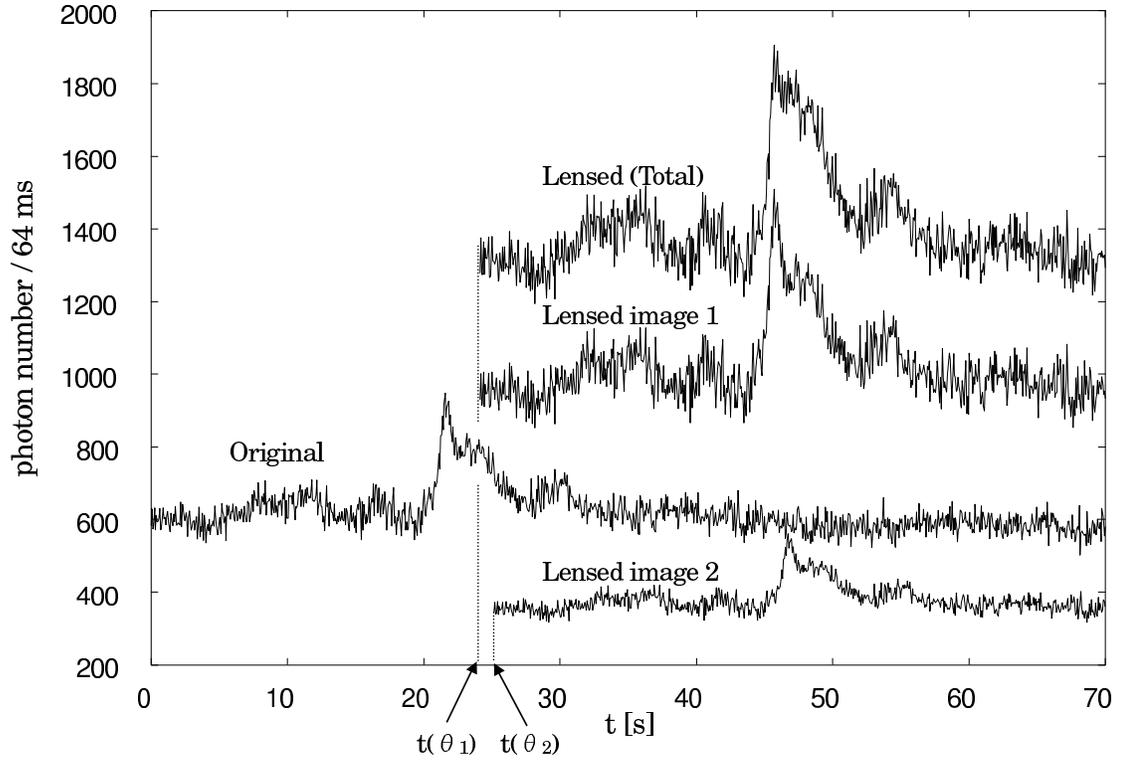}
\caption{The original light curve of GRB~930214 obtained
by {\it BATSE} and artificially lensed light curves are presented. 
The lens is assumed to be a Pop III star with $M_{\rm{L}}=10^3M_{\odot}$
at $z_{\rm{L}}=50$, while the source is located at $z_{\rm{S}}=51$.
Lensed image 1, image 2, and the total light curves
are shown. The time delay between two images is $\Delta
t = 1.0$ s 
under the assumption of $f = 0.5$.}
\label{fig:art_lens}
\end{center}
\end{figure}

However, the actual GRB light curves generally exhibit 
variabilities with time-scale shorter than $\Delta t$. 
Thus, it is not straightforward
to extract the time delay information.
To demonstrate this difficulty,
we show the light curve of GRB~930214 and
the artificially lensed light curve in Figure \ref{fig:art_lens}, where 
$M_{\rm{L}}=10^3M_{\odot}$, $z_{\rm{L}}=50$, and $f = 0.5$ are assumed. 
The time delay is $\Delta t = 1.0$ s in this case. This figure clearly shows
that if we observe only the superimposed lensed light curve, it seems 
impossible to recognize by appearance that this light curve is lensed.
Hence, we invoke a new technique to discriminate a lensed GRB
from unlensed one. 

\section{Autocorrelation Method}

\subsection{Theory}

\begin{figure} [p]
\begin{center}
\includegraphics[width=15cm,clip]{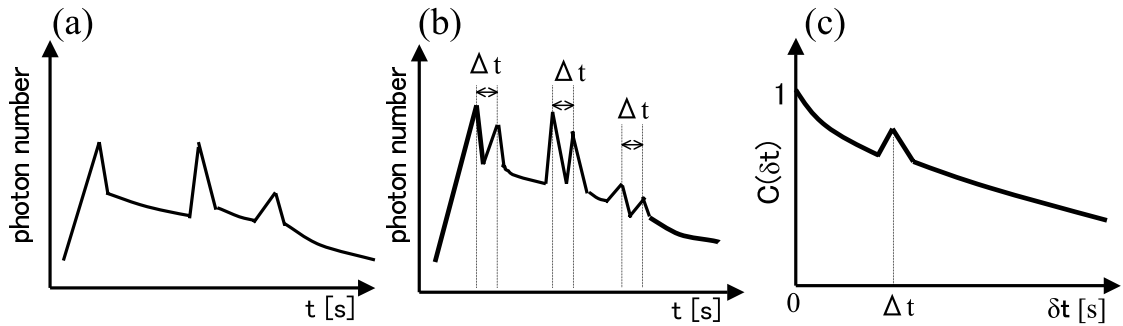}
\caption{(a) Schematic figure of a spiky GRB light curve. 
(b) The light curve as the superposition of two lensed light curves 
with the time delay of $\Delta t$. 
(c) Autocorrelation function calculated for light curve (b).}
\label{fig:correaltion}
\end{center}
\end{figure}

We pay attention to the fact that all spikes in a light
curve are individually lensed. Then, many pairs with time 
separation of $\Delta t$ appear in the light curve, as 
schematically shown in Figure \ref{fig:correaltion} (b). 
To detect those pairs, we employ the autocorrelation method 
(e.g., Geiger \& Schneider 1996). 
The autocorrelation, $C(\delta t)$, is defined as
\begin{equation}
\displaystyle C(\delta t)=\frac{\sum_{i} {I(t_{i}+\delta
t)I(t_{i})}}{\sum_{i} {I(t_{i})^2}} \label{eq:corre},
\end{equation}
where $I(t_{i})$ is the number of photons contained in
a {\it i}-th bin in the GRB light curve. 
If there are pairs with $\Delta t$, the autocorrelation
(\ref{eq:corre}) is expected to show the resonance ``bump'' around
$\Delta t$, as shown in Figure \ref{fig:correaltion} (c). 
Then, we can evaluate the time delay by the existence of this bump. 

\subsection{Robustness}

The autocorrelation method is simple and well defined, but
the issue we should check is its applicability for the actual
GRB light curves. To test the robustness of this method,
we produce artificially lensed light curves for GRBs in {\it BATSE}
archived data, and calculate the autocorrelation. 
We use 1821 light curves in the {\it BATSE} catalogue 
with the time resolution of 64 ms
\footnote{$\rm{http://cossc.gsfc.nasa.gov/batse/BATSE\_Ctlg/duration.html}$}.
Unless otherwise specified, we adopt the data of $T_{90}$,
where $T_{90}$ is defined by the duration such that
the cumulative photon counts increase from
5\% to 95\% of the total GRB photon counts (Kouveliotou
et. al. 1993, Koshut et. al. 1996). 
Then, the summation in equation (\ref{eq:corre}) is taken 
in the range of $T_{90}- \delta t$.
But, if the data in $T_{90}$ start with the bins whose
time resolution is worse than 1024 ms, we neglect those low resolution bins.

\begin{figure} [p]
\begin{center}
\includegraphics[width=12cm,clip]{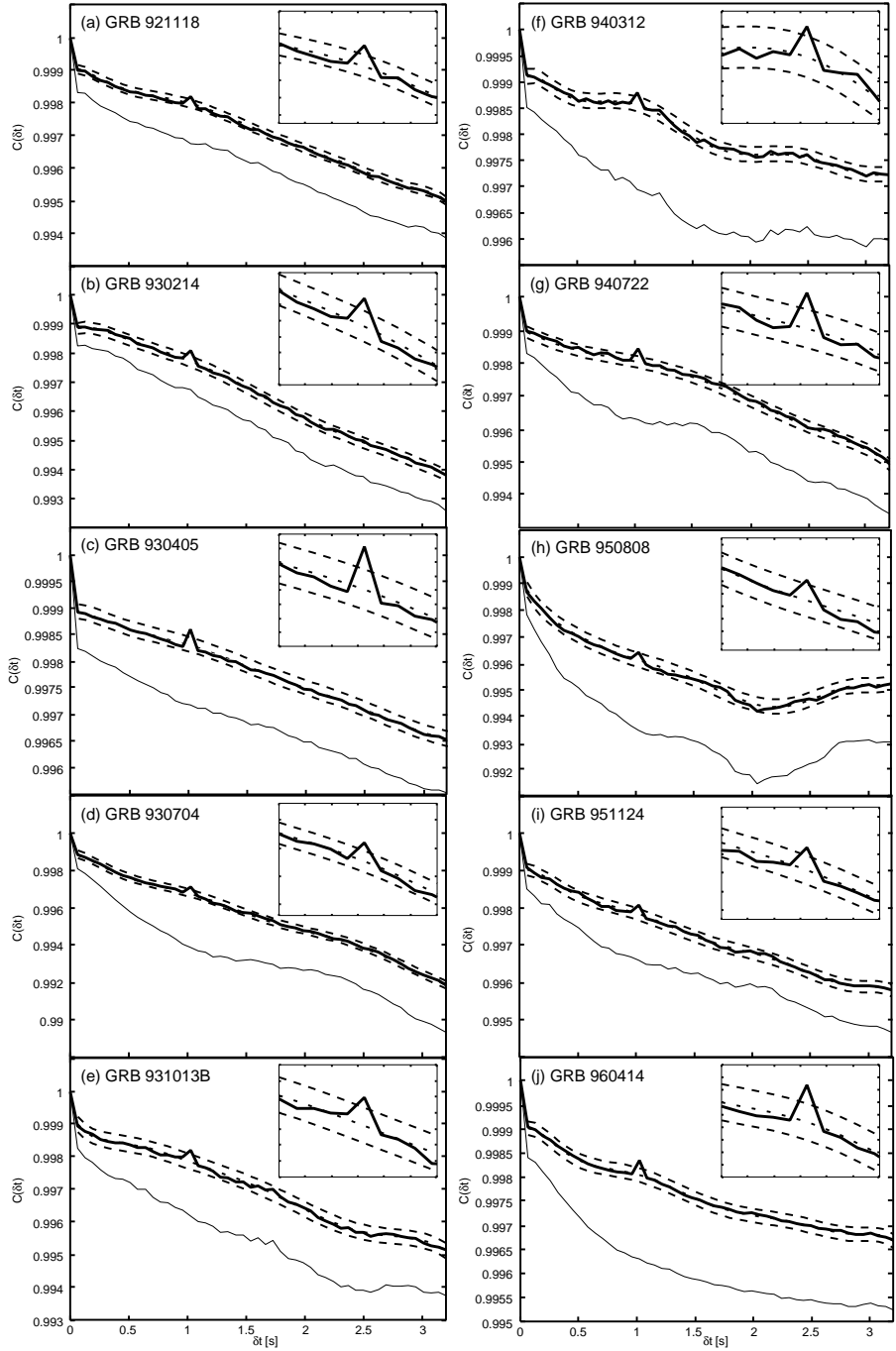}
\caption{ Autocorrelation against 
$\delta t$ for 10 GRB light
curves. In each panel, a thin curve shows the autocorrelation
for the original light curve, while a thick curve is that
for the artificially lensed light curve. Here, 
$M_{\rm{L}}=10^3M_{\odot}$, $z_{\rm{L}}=50$, and 
$f = 0.5$ are assumed and therefore the time delay 
of lensed images is $\Delta t =1$ s.
A dotted curve is the best fitting for the autocorrelation for lensed light curve, 
and two dashed curves show $\pm 3\sigma$ level from the best fitting.
In each panel, a zoomed view around a correlation bump is also shown.}
\label{fig:art_corre}
\end{center}
\end{figure}

In Figure \ref{fig:art_corre}, the resultant autocorrelation is shown
for 10 GRB light curves. 
In each panel, a thin solid line represents the autocorrelation
for the original light curve, while a thick solid line is
the autocorrelation for the artificially lensed light curve, 
where $M_{\rm{L}}=10^3M_{\odot}$, $z_{\rm{L}}=50$, and 
$f = 0.5$ are assumed, the same as Figure \ref{fig:art_lens}, 
and therefore the time delay of lensed images is $\Delta
t =1$ s. 
We can see that there is no bump in $C(\delta t)$ for
the original light curve, whereas a bump emerges
around $\delta t =1$ s in $C(\delta t)$ for the
artificially lensed light curve. Note that $C(\delta t)$
for the artificially lensed light curve is stronger than $C(\delta t)$ 
for all of the original light curve, owing to the amplification by
gravitational lensing. 
$C(\delta t)$ for the artificially lensed light curve is
fit by the polynomial of 8th degree. 
With using the best fit polynomial $F(\delta t)$,
we define the dispersion, $\sigma$, of the autocorrelation curve 
by ${\sigma}^2 =
\sum_{j=1}^{n} [C({\delta t}_j)
-F({\delta t}_j)]^2/n$, where $n$ is the
number of bins. The levels of $\pm 3\sigma $ are shown by
dashed lines. 
The zoomed view around the bump of $C(\delta t)$
for the artificially lensed light curve is also shown 
in each small panel. 
For these GRBs, bumps exceeding $3\sigma$ 
appear if lensed, corresponding to the time delay between
two lensed images, $\Delta t$.

\subsection{Dependence on $f$}
\label{f-probability} 

\begin{figure} [p]
\begin{center}
\includegraphics[width=15cm,clip]{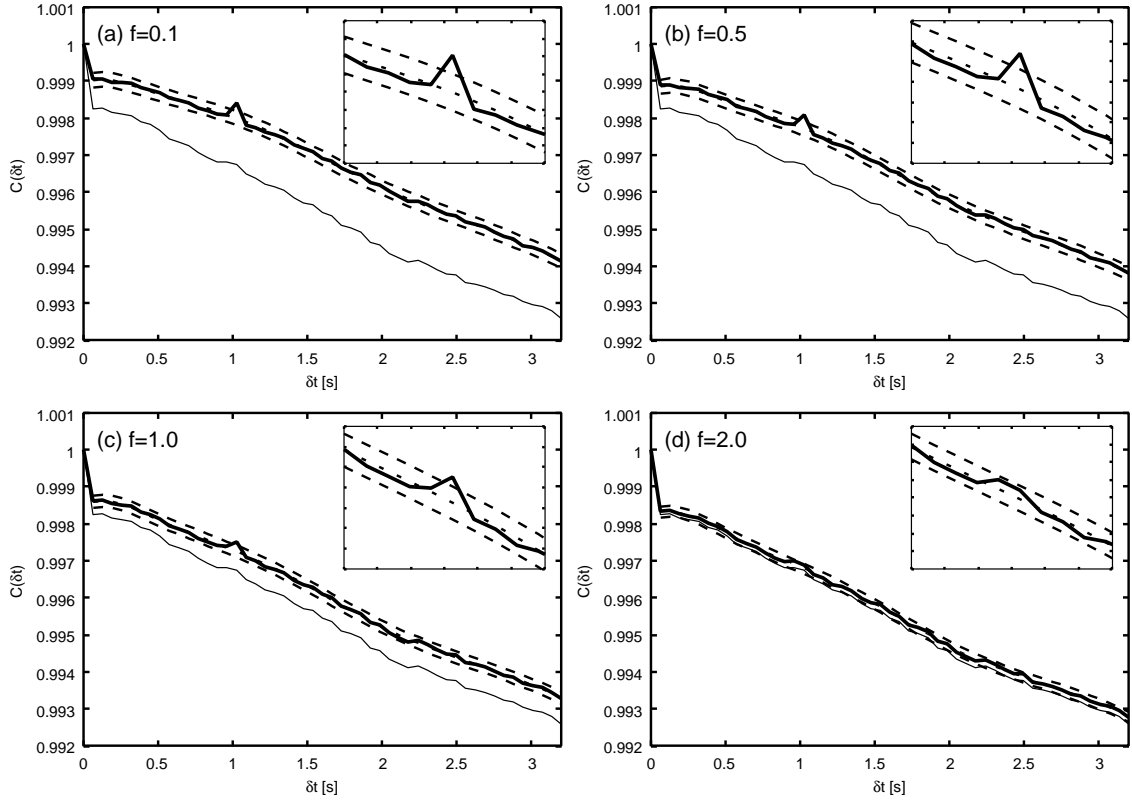}
\caption{Dependence on $f=\beta/\theta_{\rm E}$ of the autocorrelation for
the artificially lensed light curve of GRB~930214, which is the same
GRB as panel (b) in Fig. \ref{fig:art_corre}.
The time delay is $\Delta t=0.1$ s. The meanings of curves
are the same as Fig. \ref{fig:art_corre}.
It is seen that the bump is weaker for larger $f$. 
}
\label{fig:f-dependence}
\end{center}
\end{figure}

\begin{figure} [p] 
\begin{center}
\includegraphics[width=15cm,clip]{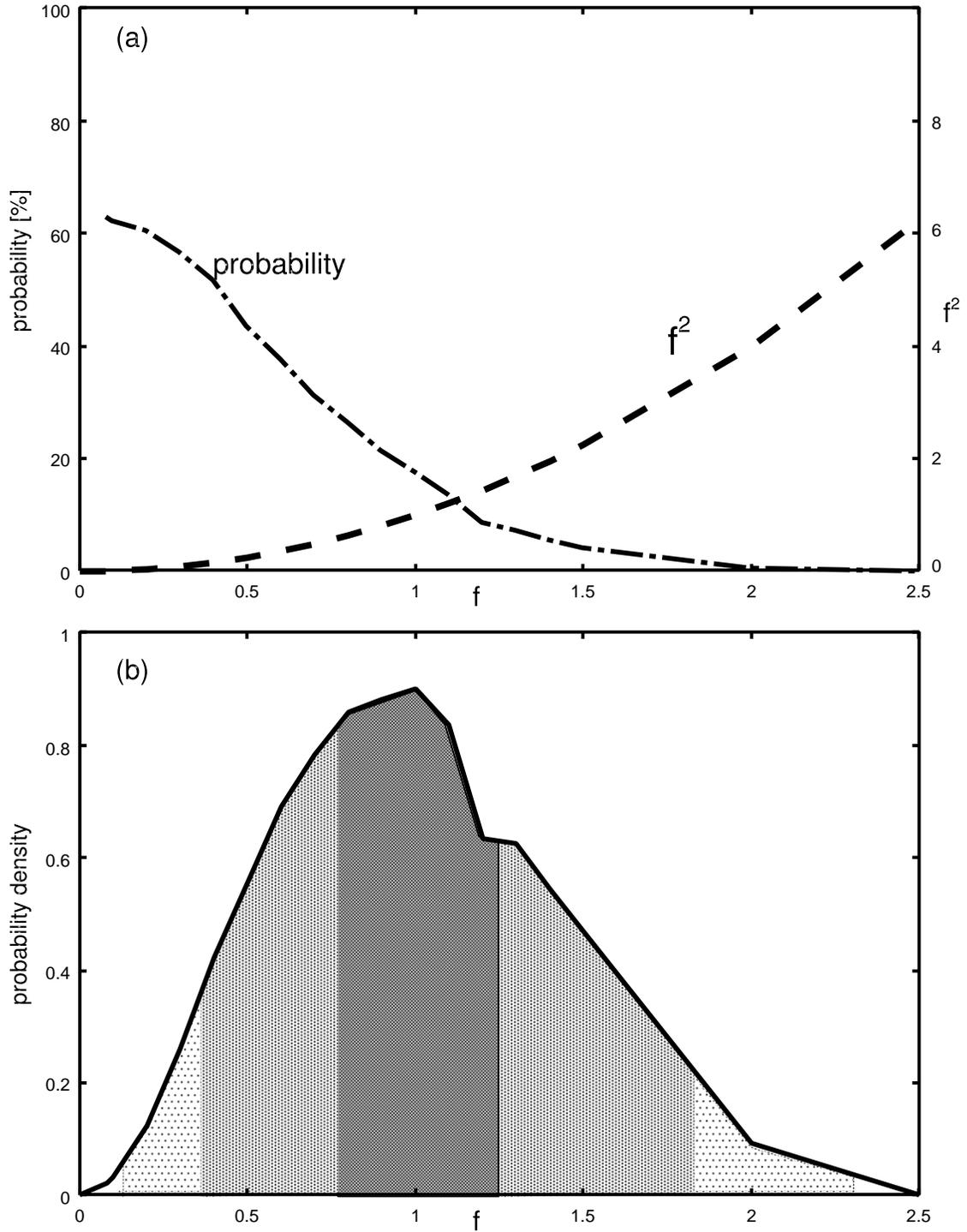}
\caption{(a) A dot-dashed curve shows the probability 
of exhibiting bumps exceeding $3\sigma$. Here,
$\Delta t=1.0 \ \rm{s}$ is assumed.
A dashed line is $f^2$, which is proportional to 
the cross-section of gravitational lens.
(b) The probability density of exhibiting bumps against $f$. 
The gray scales represent $1\sigma$, $2\sigma$, and $3\sigma$
around $f=1$ (from dark to light).}
\label{fig:prob-dens}
\end{center}
\end{figure}

In fact, not all GRB light curves exhibit bumps in 
$C(\delta t)$ when lensed. 
The fraction of GRBs which show bumps exceeding $3\sigma$
in $C(\delta t)$ depends on the value of $f=\beta/\theta_{\rm E}$.
Also, the time delay $\Delta t$ and the
magnification $A_{1,2}$ depend on $f$. 
To demonstrate this, we show the dependence on $f$ of 
the autocorrelation for the artificially lensed light curve of
GRB~930214, in Figure \ref{fig:f-dependence}.
It is clear that
if $f$ is larger, the bump is suppressed.
This is because the contribution of image 2 becomes
smaller with increasing $f$. 
In this GRB, the bump over $3\sigma$ disappears at $f > 1.0$. 
The value of $f$ at which the bump disappears differs in each GRB. 
Using 220 GRB data, which are used in 
Fenimore \& Ramirez-Ruiz (2000), 
we obtain the fraction of GRBs which show bumps over $3\sigma$
in $C(\delta t)$ as a function of $f$. In this analysis, 
$\Delta t= 1.0 $ s is assumed. 
The resultant fraction is shown in Figure \ref{fig:prob-dens} (a).
For $f=0.5$, about a half of GRBs show bumps exceeding $3\sigma$
in $C(\delta t)$. In contrast, the cross-section of
the gravitational lensing is proportional to $f^2$, which is also
shown in Figure \ref{fig:prob-dens} (a). 
We evaluate the probability density of GRBs exhibiting bumps 
over $3\sigma$ by multiplying the fraction for which a bump appears by $f^2$. 
The normalized probability density against $f$ is shown in 
Figure \ref{fig:prob-dens} (b). As a result,
the probability is peaked around $f=1$ and the standard deviation 
corresponds to $\Delta f \approx 0.25$.
It is noted that this probability density is found to be hardly dependent 
on the value of $\Delta t$.

\subsection{Optical depth}

\begin{figure} [p]
\begin{center}
\includegraphics[width=15cm,clip]{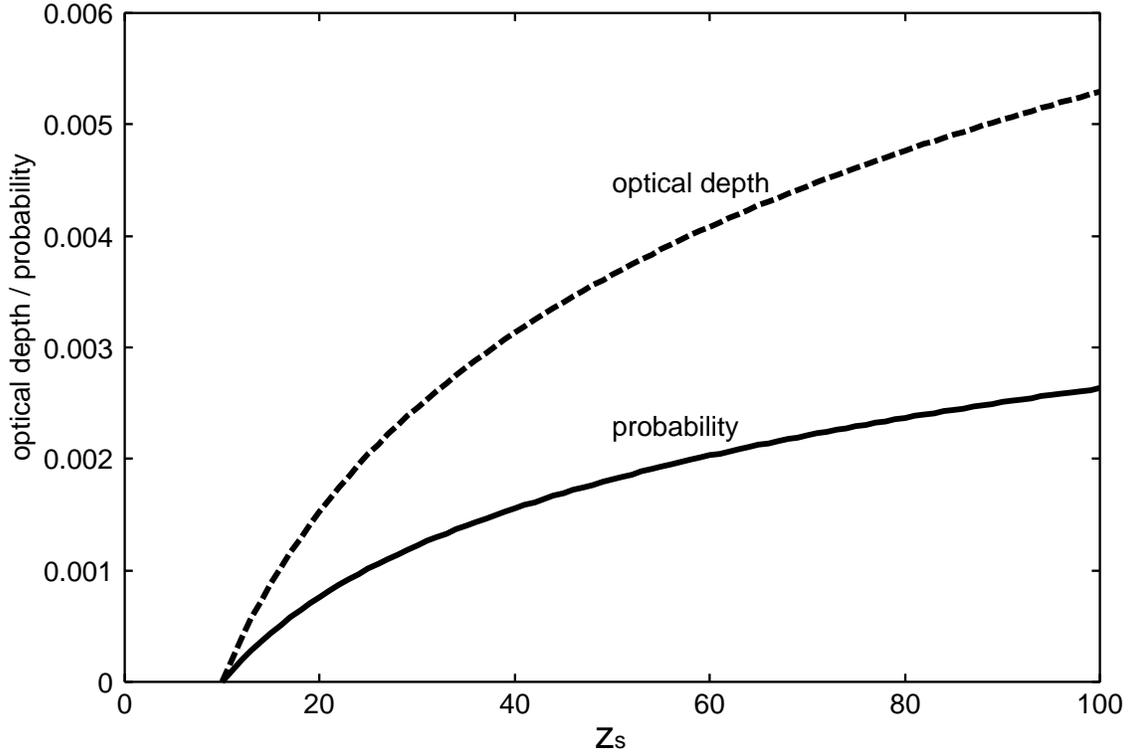}
\caption{ Probability of bump detection for lensed GRBs, 
which is shown by a solid line.
A dashed line is the optical depth of gravitational lensing,
when a GRB as a source object is located at $z_{\rm{S}} \geq 10$
and 10\% of $\Omega_{\rm{b}}$ at $z_{\rm{L}}\geq 10$ 
contributes to gravitational lensing as a lens object. 
}
\label{fig:optical-depth}
\end{center}
\end{figure}

Here, we estimate the optical depth of gravitational lensing 
by Pop III stars.
If we assume that the fraction $\alpha$ of baryonic matter
composes Pop III stars at $z \geq z_{\rm{III}}$,
then the optical depth is given by 
\begin{eqnarray}
\displaystyle \tau(z_{\rm{S}})
& = & \int_{z_{\rm{III}}}^{{z}_{\rm{S}}} n_{\rm{L}}(z_{\rm{L}}) 
\sigma_{\rm{L}} \frac{c dt}{dz_{\rm{L}}} dz_{\rm{L}} 
\label{eq:optical},
\end{eqnarray}
where
$\sigma_{\rm{L}}$ is the cross-section given as $\sigma_{\rm{L}}=\pi
(D_{\rm{OL}}\theta_{\rm{E}})^2$ and $n_{\rm{L}}(z_{\rm{L}})$ is
the number density of the lens objects given as
\begin{equation}
 n_{\rm{L}}(z_{\rm{L}}) = \alpha \frac{3H_0^2}{8\pi G} 
  \frac{\Omega_{\rm{b}}}{M_{\rm{L}}} (1+z_{\rm{L}})^3.
\end{equation}
Then, the optical depth (\ref{eq:optical}) is calculated as 
\begin{eqnarray}
\displaystyle \tau(z_{\rm{S}})
& = &  \frac{3}{5}\alpha \Omega_{\rm{b}}
\left\{\left[\frac{(1+z_{\rm{S}})^{5/2}+1}
{(1+z_{\rm{S}})^{5/2}-1}\right]\ln(1+z_{\rm{S}}) \right. \nonumber \\
& &  \left. \hspace*{0.8cm} -\left[\frac{(1+z_{\rm{III}})^{5/2}+1}
{(1+z_{\rm{III}})^{5/2}-1}\right]\ln(1+z_{\rm{III}}) \right\}
\end{eqnarray}
in the Einstein-de Sitter universe
(Turner, Ostriker, \& Gott 1984; Turner \& Umemura 1997).
Here, we assume $\alpha=0.1$ and $z_{\rm{III}}=10$. 
The resultant optical depth of
Pop III star lensing is shown in Fig. \ref{fig:optical-depth}.
Since $\tau(z_{\rm{S}})$ is the probability that a source is located
inside the Einstein ring radius ($f \leq 1$), 
it is not the probability of bump detection.
The probability of bump appearance, $p(f)$, 
is a decreasing function of $f$,
as shown in Fig. \ref{fig:prob-dens} (a).
Since the optical depth that $f$ is in the range of
$[f,f+df]$ is given by 
$d\tau(z_{\rm{S}})=\tau(z_{\rm{S}})df^2 = \tau(z_{\rm{S}}) 2fdf$,
the probability of bump detection is given by
\begin{equation}
\displaystyle P(z_{\rm{S}})=\int_0^{\infty} 2f p(f)  
\tau(z_{\rm{S}}) df \label{eq:prob}.
\end{equation}
The resultant bump detection probability is also
shown in Fig. \ref{fig:optical-depth}.
From this figure, the probability turns out to be 
$\approx 0.001$ for $z_{\rm{S}}=20-40$. 
If we take into account that more than 10-30\% of GRBs 
occur at $z \gtrsim 10$ (Bromm \& Loeb 2002), 
the expectation number of bump detection for lensed GRBs 
is assessed to be one in a few thousand GRBs.

\section{A Candidate for GRB Lensed at $z \approx 60$}

\subsection{Data analysis}

\begin{figure} [p]
\begin{center}
\includegraphics[width=15cm,clip]{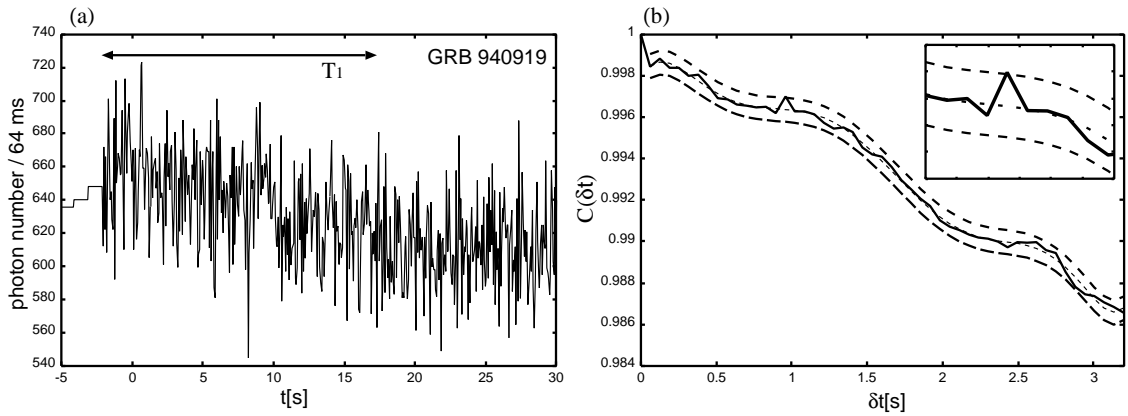}
\caption{(a) The intrinsic light curve of GRB~940919. 
(b) Autocorrelation against $\delta t$. A $3\sigma$ bump
appears at $\Delta t=0.96$ s. }
\label{fig:GRB940919-corre}
\end{center}
\end{figure}

As shown above, the autocorrelation of intrinsic light
curves exhibits no bumps for almost all GRBs. But, 
a few in 1000 GRBs might show bumps in $C(\delta t)$
even for intrinsic light curves. Hence, 
we calculate the autocorrelation of all GRB light curves 
available in {\it BATSE} catalogue, which amount to 1821 GRBs. 
As a result, we have found one candidate, 
GRB~940919 ({\it BATSE} trigger number 3174), in which a $3\sigma$ bump 
in $C(\delta t)$ appears.
The light curve and the autocorrelation of this GRB
is shown in Figure \ref{fig:GRB940919-corre}. 
As seen in panel (b), a bump exceeding $3\sigma$
appears at $\Delta t=0.96$ s.

\subsection{Statistical significance}

To check the statistical significance of a bump in $C(\delta t)$ 
of GRB~940919, we make a test with mock light curves.
Here, we generate mock light curves using a smoothed correlation
function that dose not show any bump, and investigate whether
bumps appear in correlation functions just from pure statistical 
fluctuations. 

\begin{figure} [p] 
\begin{center}
\includegraphics[width=10cm,clip]{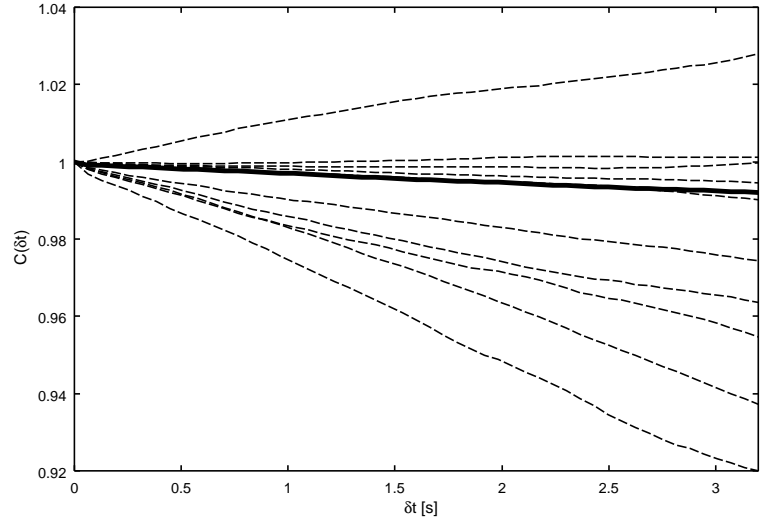}
\caption{ Autocorrelation for mock light
curves. A thick curve is the autocorrelation for the
original light curve (GRB~000421). Other thin dashed curves are
a part of autocorrelations for 2000 mock light curves
based on the same GRB. }
\label{fig:re_corre} 
\end{center}
\end{figure}

From Wiener-Kihntchine theorem, the power
spectrum of light curves are given by
\begin{equation}
\textstyle  |I(\omega)|^2 = \mathcal{F} \left[ \sum_{i} I (t_{\rm{i}}) I
(t_{\rm{i}} + \delta t) \right]
\label{eq:wiener-khintchine},
\end{equation} 
where $\mathcal{F}$ denotes the Fourier
transformation and $\omega = 2 \pi / \delta t$. 
We can generate mock light curves by the inverse Fourier
transformation of $I(\omega)$. Here, in order to add fluctuation 
to $I(\omega)$, we take random Gaussian distributions, 
where $|I(\omega)|$ is the standard deviation and 
the phase is random in the range of $[ 0,2\pi ]$.
Then, the mock light curve is given by
\begin{equation}
\textstyle \tilde{I}(t) = \mathcal{F}^{-1}[I(\omega)]= 
\sum_{\omega} |\tilde{I}(\omega)| \cos (-\phi - \omega t)
\label{eq:inv-fourier} ,
\end{equation} 
where $\tilde{I}(\omega)$ is a random sample in the Gaussian
distribution, and $\phi$ is the random phase shift from 0 to 2$\pi$.
We produce 2000 mock light curves using a correlation
function, and recalculate the autocorrelation $C(\delta t)$ 
by equation (\ref{eq:corre}). As a result, we have found that
no bump higher than $3 \sigma$ appears in the $C(\delta t)$ 
of 2000 mock light curves.
A part of the recalculated $C(\delta t)$ are shown in Figure \ref{fig:re_corre}. 
Thus, it is unlikely that a bump in the correlation arises as a result of 
pure statistical fluctuations. 

\subsection{Light curve decomposition}

\begin{figure} [p] 
\begin{center}
\includegraphics[width=10cm,clip]{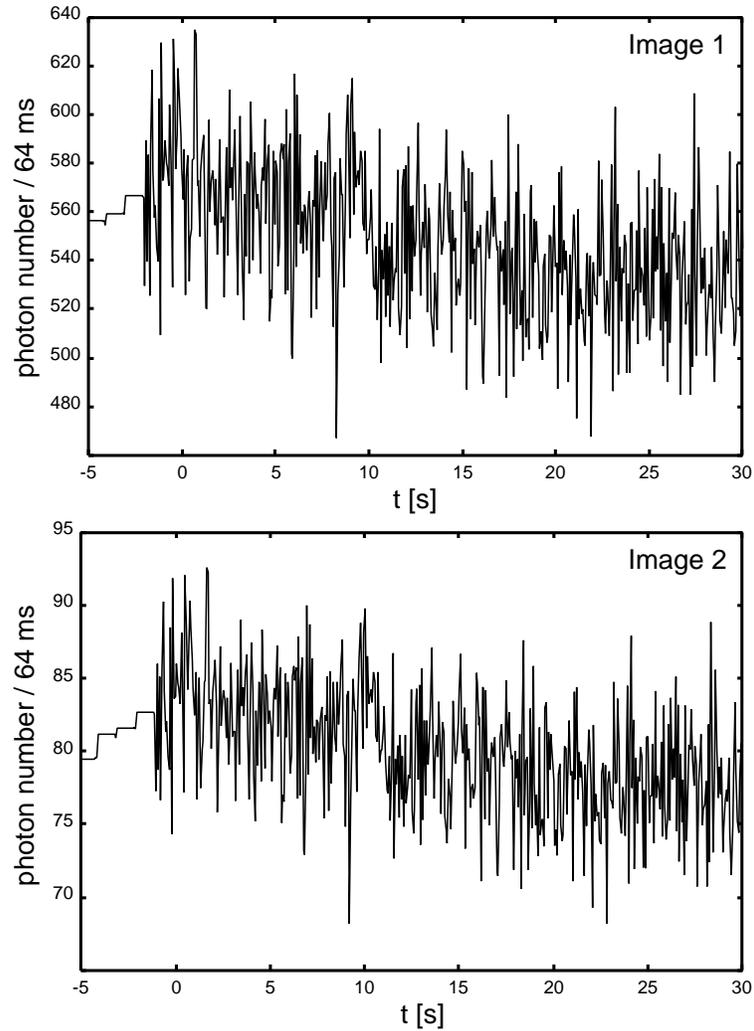}
\caption{The decomposed light curves for GRB~940919, assuming
that the observed light curve is the superposition of two lensed light 
curves with $\Delta t=0.96$ s.} 
\label{fig:decomp}
\end{center}
\end{figure}

\begin{figure} [p] 
\begin{center}
\includegraphics[width=10cm,clip]{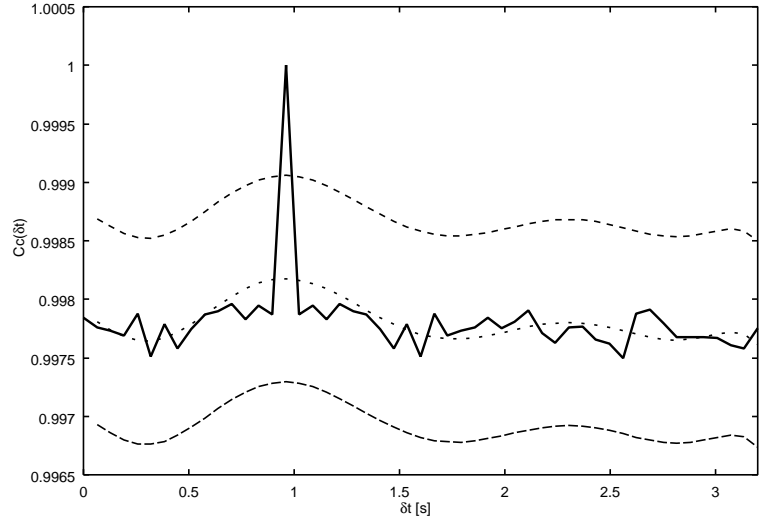}
\caption{The cross-correlation of decomposed light curves 
shown in Fig. \ref{fig:decomp}. A dotted line is the best fit
curve, and short-dashed curves are $\pm 3 \sigma$ levels.
The cross-correlation shows a peak well above $3 \sigma$,
when $\delta t$ is the same as $\Delta t =0.96$ s. 
}
\label{fig:cross-corr}
\end{center}
\end{figure}

As a further test for the lensing of GRB~940919 light curve,
we attempt to decompose the light curve, assuming
that it is the superposition of two lensed light curves 
with $\Delta t=0.96$ s, and analyze the decomposed light curves.
The decomposition is made by the following recurrence formula; 
\begin{equation}
\displaystyle
I_{\rm{tot}} (t) = I_1 (t) + I_2 (t),  \label{eq:total} 
\end{equation}
\begin{equation}
\displaystyle
I_2 (t) = \frac{A_2}{A_1} I_1 (t - \Delta t), \label{eq:image2} 
\end{equation}
where $I_{\rm{tot}} (t)$ is the observed intensity, and 
$I_1 (t)$ and $I_2 (t)$ are intensities for image 1 and 2,
respectively. If these two equations are combined,
$I_1 (t)$ can be expressed by 
\begin{equation}
I_1 (t) = 
\sum_{j=0}^N \left(- \frac{A_2}{A_1}\right)^j
I_{\rm{tot}} (t - j \Delta t). 
\label{eq:separate} 
\end{equation}
The summation is taken in the range of $t - j \Delta t \geqq 0$, where
$t = 0$ is the starting point of $T_{90}$. Then, we can
derive also $I_2 (t)$ by equation (\ref{eq:total}). 
In Figure \ref{fig:decomp}, 
the decomposed light curves are shown in the case of $f = 1$ which is
the most probable case as shown in \S \ref{f-probability}.
The application of this decomposition method for the finite amount of data
does not guarantee that the light curve is successfully decomposed
into two lensed light curves. 
Hence, to check the validity of this decomposition method, we calculate the 
cross-correlation of two decomposed light curves by
\begin{equation}
\displaystyle C_{\rm{c}}(\delta t) = \frac{\sum_{i}
{I_1(t_{i}) I_2(t_{i} + \delta t)}}{\sqrt{\sum_{i}
{I_1(t_{i})^2}} \sqrt{\sum_{i} {I_2(t_{i} + \delta t)^2} } },  \label{eq:cross_corre}
\end{equation}
where $\delta t$ is the time shift. 
The result is shown in Figure \ref{fig:cross-corr}.
As clearly shown in this figure, the cross-correlation is peaked
when $\delta t$ accords with $\Delta t=0.96$ s. 
Also, each decomposed light curve shows no bump
higher than $3\sigma$ in the autocorrelation for a reasonable range of $f$.
Hence, we can conclude that the light curve of GRB~940919 
is successfully decomposed and is likely to be
the superposition of two lensed light curves. 

\subsection{Redshift estimation}

\begin{figure} [p] 
\begin{center}
\includegraphics[width=15cm,clip]{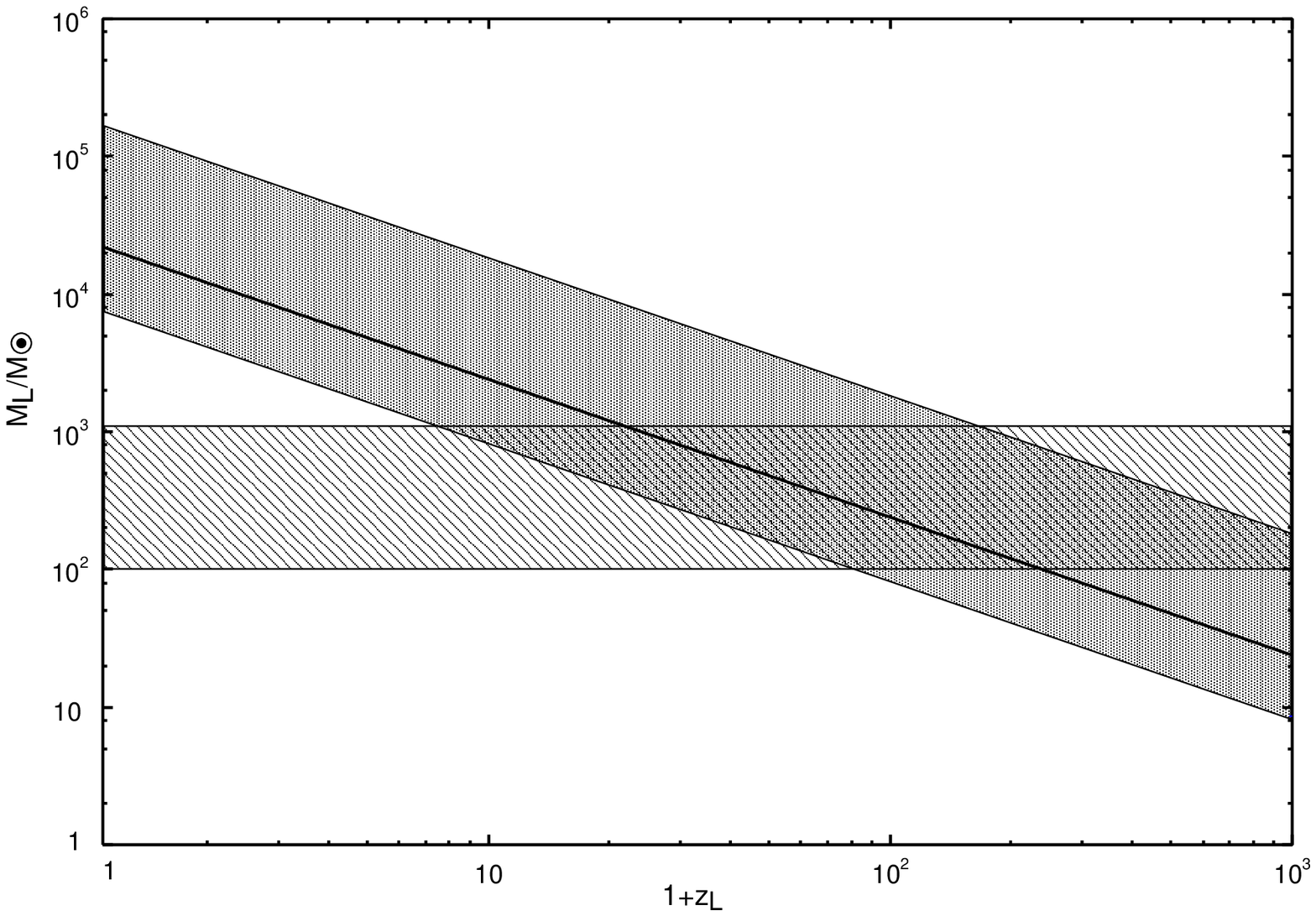}
\caption{The dark gray region shows the suitable range of
$M_{\rm{L}}(1+\rm{z_{\rm{L}}}) $ for GRB~940919.
Here, $3\sigma$ range of $f$ (see Fig. \ref{fig:prob-dens}b) 
is adopted. The central thick solid line corresponds to
the case of $f=1$, which is most probable. 
The hatched region is the mass range of Pop III stars
by Nakamura \& Umemura (2001).}
\label{fig:z-range}
\end{center}
\end{figure}

Here, we constrain the redshift of the lens
object. As shown in equation (\ref{eq:delta}),
$\Delta t$ just determines $M_{\rm{L}}(1+z_{\rm{L}})$,
except for $f$.
If the probability density against $f$ (Fig. \ref{fig:prob-dens} (b) ) 
is applied, we can derive a suitable range for $M_{\rm{L}}(1+z_{\rm{L}})$.
The dark gray region in Figure \ref{fig:z-range} represents 
the suitable range for $\Delta t=0.96$ s. 
On the other hand, the hatched region shows the mass range 
of Pop III stars obtained by Nakamura \& Umemura (2001). 
Combining these two regions, the allowed redshift of the lens object 
is at least $ z_{\rm{L}} \approx 10$.
If $f=1$ is adopted, the redshift ranges from 
$ z_{\rm{L}} \approx 20$ to $ z_{\rm{L}} \approx 200$, 
where the most probable one is $z_{\rm{L}} \approx 60$.

Nonetheless, there still exists another possibility
that the lens object is as massive as $\sim 10^4 M_{\odot}$ 
located at $z_{\rm{L}}\simeq$0, as seen in
Fig. \ref{fig:z-range}. Loeb (1993) and Umemura et al. (1993)
suggested that relic massive black holes are candidates
for such an object.
Sasaki \& Umemura (1996) 
place a constraint on $\Omega_{\rm{BH}}$ from the UV background 
intensity and the Gunn-Peterson effect in the context 
of a cold dark matter cosmology. They find that the black hole 
mass density might be as low as 
$\Omega_{\rm{BH}}/\Omega_{\rm{b}} \lesssim 10^{-3}$.
Therefore, the expected number of bump detection for
massive black holes is by two orders of magnitude lower
than that for Pop III stars.  

\section{Conclusions}   

To place constraints on the redshifts of GRBs which originate from
Pop III stars at $z>10$, we have proposed a novel method 
based on the gravitational lensing effects. 
If the lens is Pop III stars with $10^2-10^3M_\odot$ at $z>10$, 
the time delay between two lensed images of a GRB
is $\approx 1$ s. This time delay is longer than the time
resolution (64 ms) of GRB light curves and shorter than the
duration of GRB events. Therefore, if a GRB is lensed,
we observe the superposition of two lensed light curves. 
We have considered the autocorrelation method 
to extract the imprint of 
gravitational lensing by Pop III stars in the GRB light curves.
Using {\it BATSE} data, we have derived the probability of the resonance bump in
the autocorrelation function, which is an indicator for the gravitational lensing. 
Applying this autocorrelation method to GRB light curves 
in the {\it BATSE} catalogue, 
we have demonstrated that more than half of the light curves can
show resonance bumps, if they are lensed. 
Furthermore, in 1821 GRB light curves,
we have found one candidate of GRB lensed by a Pop III star at $z \approx 60$.
The present method is quite straightforward and therefore provides 
an effective tool to search for Pop III stars at redshift greater than 10.
Although the number of GRBs with available data is 1821 in this paper, 
the {\it Swift} satellite is now accumulating more GRB data. 
If the present method is applied for those data, 
more candidates for GRBs lensed at $z>10$ may be found in the future. 
These can provide a firm evidence of massive Pop III stars born at
high redshifts. 

\acknowledgments
We thank T. Murakami and D. Yonetoku for helpful information 
and fruitful discussion, and S.Mao for his valuable comments.
Numerical simulations were performed 
with facilities at the Center for Computational Sciences,
University of Tsukuba. 
One of the author (AY) acknowledges to 
the Japan Society for the Promotion of Science 09514, 
Inoue Foundation for Science, and 
JSPS Postdoctoral Fellowships for Research Abroad.   
This work was supported in part 
by Grants-in-Aid for Scientific Research from MEXT 
16002003 (MU).

\clearpage

\begin{thebibliography}{}
\bibitem[]{} Amati, L., et al. 2002, A\&A, 390, 81
\bibitem[]{} Andersen, M. I., et al. 2000, A\&A, 364,
L54
\bibitem[]{} Baltz, E. A., \& Hui, L. 2005, \apj, 618, 403
\bibitem[]{} Blaes, O. M., \& Webster, R. L. 1992, \apj, 391, L63
\bibitem[]{} Bloom, J. S., Frail, D. A., \& Kulkarni, S. R. 2003, \apj, 594, 674
\bibitem[]{} Bromm, V., \& Loeb, A. 2002 \apj, 575, 111
\bibitem[]{} Fenimore, E. E., \& Ramirez-Ruiz, E. 2000,
astro-ph/0004176
\bibitem[]{} Garnavich, P. M., Loeb, A., \& Stanek,
K. Z. 2000, \apj, 544, L11
\bibitem[]{ } Gehrels, N., et al.\ 2004, \apj, 611, 1005
\bibitem[]{} Geiger, B., \& Schneider, P. 1996, MNRAS,
282, 530
\bibitem[]{} Heger, A., \& Woosley, S. E. 2002, \apj, 567, 532
\bibitem[]{} Heger, A., et al. 2003, \apj, 591, 288
\bibitem[]{} Kawabata, K. S., et al. 2003, \apj, 593,
L19
\bibitem[]{} Kawai, N. et al. 2006, Nature, 440, 184 
\bibitem[]{} Kogut, A., et al. 2003, \apjs, 148, 161 
\bibitem[]{} Koopmans, L. V. E., \& Wambsganss, J. 2001,
MNRAS, 325, 1317 
\bibitem[]{} Koshut, T. M., Paciesas, W. S., Kouveliotou, C., van Paradijs, J., Pendleton, G. N., Fishman, G.J., \& Meegan, C. A. 1996, \apj, 463, 570
\bibitem[]{} Kouveliotou, C., et al. 1993, \apj, 413, L101
\bibitem[]{} Lamb, D. Q., \& Reichart, D. E. 2000, \apj, 536, 1
\bibitem[]{} Lloyd-Ronning, N. M, Fryer, C. L., \& Ramirez-Ruiz, E. 2002, \apj, 574, 554
\bibitem[]{} Loeb, A. 1993, \apj, 403, 542
\bibitem[]{} Loeb, A., \& Perna, R. 1998, \apj, 495, 597
\bibitem[]{} Marni G. F., Nemiroff, R. J., Norris,
J. P., Hurley, K., \& Bonnell, J. T. 1999, \apj, 512, L13
\bibitem[]{} Metzger, M., et al. 1997, Nature, 387, 879
\bibitem[]{} Mizusawa, H., Nishi, R., \& Omukai, K.\ 2004, \pasj, 56, 487 
\bibitem[]{} Murakami, T., Yonetoku, D., Umemura, M.,
Matsubayashi, T. Yamazaki, R. 2005, \apj, 625, L13
\bibitem[]{} Nakamura, F., \& Umemura, M., 2001, \apj, 548, 19
\bibitem[]{} Narayan, R., \& Bartelmann, M. 1999, in
Formation of Structure in the Universe, ed. A. Dekel \&
J. P. Ostriker (Cambridge: Cambridge Univ. Press), 360
\bibitem[]{} Nemiroff, R. J., et al. 1993, \apj, 414, 36
\bibitem[]{} Nemiroff, R. J., \& Marani, G.F. 1998, \apj, 494, L173
\bibitem[]{} Norris, J. P., Marani, G. F., \& Bonnell, J. T. 2000, \apj, 534, 248
\bibitem[]{} Paczy\'nski, B. 1986, \apj, 308, L43
\bibitem[]{} Paczy\'nski, B. 1987, \apj, 317, 51
\bibitem[]{} Page, L. et al. 2006, astro-ph/0603450
\bibitem[]{} Price, P. A., et al 2003, Nature, 423, 844
\bibitem[]{} Sasaki, S., \& Umemura, M. 1996, \apj, 462, 104
\bibitem[]{} Schaefer, B. E., Deng, M., \& Band D. L. 2001, \apj, 563, L123
\bibitem[]{} Spergel, D., et al. 2003, \apjs, 148, 175
\bibitem[]{} Spergel, D., et al. 2006, astro-ph/0603449
\bibitem[]{} Turner, E. L., Ostriker, J. P., \& Gott, J. R. 1984, \apj, 284, 1
\bibitem[]{} Turner, E. L., \& Umemura, M. 1997, \apj, 483, 603
\bibitem[]{} Uemura, M., et al. 2003, Nature, 423, 843
\bibitem[]{} Umemura, M., Loeb, A., \& Turner, E. L. 1993, \apj, 419, 459
\bibitem[]{} Yonetoku, D., Murakami, T., Nakamura, T., Yamazaki, R., Inoue, A., K., \& Ioka, K. 2004, \apj, 609, 935
\bibitem[]{} Williams, L. L. R., \& Wijers, R. A. M. J. 1997, MNRAS, 286, L11
\bibitem[]{} Wyithe, J. S. B., \& Turner, E. L. 2002,
\apj, 575, 650

\end{thebibliography}
\end{document}